# A Way to Understand Various Patterns of Data Mining Techniques for Selected Domains

*Dr. Kanak Saxena*
Professor & Head,
Computer Application
SATI, Vidisha,
kanak.saxena@gmail.com

*D.S. Rajpoot*
Registrar, UIT
RGPV, Bhopal
dsrphd@yahoo.com

*Abstract:* This has much in common with traditional work in statistics and machine learning. However, there are important new issues which arise because of the sheer size of the data. One of the important problem in data mining is the Classification-rule learning which involves finding rules that partition given data into predefined classes. In the data mining domain where millions of records and a large number of attributes are involved, the execution time of existing algorithms can become prohibitive, particularly in interactive applications.

## 1. Introduction :

An enormous amount of data stored in databases and data warehouses, it is increasingly important to develop [1] powerful tools for analysis of such data and mining interesting knowledge from it. Data mining [4] is a process of inferring knowledge from such huge data. It has five major components:

- Association rules

- Classification or clustering

- Characterization & Comparison

- Sequential Pattern Analysis.

- Trend Analysis

An *association rule [5]* is a rule which implies certain association relationships among a set of objects in a database. In this process we discover a set of association rules at multiple levels of abstraction from the relevant set(s) of data in a database. For example, one may discover a set of symptoms [2] often occurring together with certain kinds of diseases and further study the reasons behind them. Since finding interesting association rules in databases may disclose some useful patterns for decision support, selective marketing, financial forecast, medical diagnosis and many other applications, it has attracted [3] a lot of attention in recent data mining research. Mining association rules may require iterative scanning of large transaction or relational databases which is quite costly in processing.

## 2. A brief review of the work already done in the field :

Sequential pattern mining is an interesting data mining problem with many real-world applications. This problem has been studied extensively in static databases. However, in recent years, emerging applications have introduced a new form of data called data stream. In a data stream [6], new elements are generated continuously. This poses additional constraints on the methods used for mining such data: memory usage is restricted, the infinitely [8] flowing original dataset cannot be scanned multiple times, and current results should be available on demand. Mendes, L.F. Bolin Ding, Jiawei Han [9] introduces two effective methods for mining sequential patterns from data streams: the SS-BE method and the SS-MB method. The proposed methods break the stream into batches and only process each batch once. The two methods use different pruning strategies [10] that restrict the memory usage but can still guarantee that all true sequential patterns are output at the end of any batch. Both algorithms scale linearly in execution time as the number of sequences grows, making them effective methods for sequential pattern mining in data streams. The experimental results also show that our methods are very accurate in that only a small fraction of the patterns that are output are false positives. Even for these false positives, SS-BE guarantees that their true support is above a pre-defined threshold.

Previous studies have shown mining closed patterns provides more benefits than mining the complete set of frequent patterns, since closed pattern mining leads to more compact results and more efficient algorithms. It is quite useful in a data stream environment where memory and computation power are major concerns. Lei Chang Tengjiao Wang Dongqing Yang Hua Luan [20] studies the problem of mining closed sequential patterns over data stream sliding windows. An efficient algorithm *SeqStream* is developed to mine closed sequential patterns in stream windows incrementally, and various novel strategies are adopted in *SeqStream [7]* to prune search space aggressively. Extensive experiments on both real and synthetic data sets show that SeqStream outperforms PrefixSpan, CloSpan and BIDE by a factor of about one to two orders of magnitude.

The input data is a set of sequences, called data-sequences. Each data sequence is ordered list of transactions (or itemsets), where each transaction is a sets of items (literals). Typically there is a transaction-time associated with each transaction. A sequential pattern also consists of a list of sets of items. The problem is to find all sequential patterns with a user-specified minimum support, where the support of a sequential pattern is the percentage of data sequences that contain the pattern.

The framework of sequential pattern discovery is explained here using the example of a customer transaction database as by Agrawal & Srikant [11]. The database is a list of time-stamped transactions for each customer that visits a supermarket and the objective is to discover (temporal) buying patterns that sufficiently many customers exhibit. This is essentially an extension (by incorporation of temporal ordering information into the patterns being discovered) of the original association rule mining framework proposed for a database of unordered transaction records (Agrawal et al 1993) [12] which is known as the Apriori algorithm. Since there are many temporal pattern discovery algorithms that are modeled along the same lines as the Apriori algorithm, it is useful to first understand how Apriori works before discussing extensions to the case of temporal patterns.

Let $D$ be a database of customer transactions at a supermarket. A transaction is simply an unordered collection of items purchased by a customer in one visit to the supermarket. The Apriori algorithm [13] systematically unearths all patterns in the form of (unordered) sets of items that appear in a sizable number of transactions. We introduce some notation to precisely define this framework. A non-empty set of items is called an itemset. An itemset i is denoted by $(i_1, i_2, i_3, \cdots i_m)$, where $i_j$ is an item. Since i has m items, it is sometimes called an m-itemset. Trivially, each transaction in the database is an itemset. However, given an arbitrary itemset i, it may or may not be contained in a given transaction T. The fraction of all transactions in the database in which an itemset is contained in is called the support of that itemset. An itemset whose support exceeds a user-defined threshold is referred to as a frequent itemset. These itemsets [14] are the patterns of interest in this problem. The brute force method of determining supports for all possible itemsets (of size m for various m) is a combinatorially explosive exercise and is not feasible in large databases (which is typically the case in data mining). The problem therefore is to find an efficient algorithm to discover all frequent itemsets in the database $D$ given a user-defined minimum support threshold.

The Apriori algorithm exploits the following very simple (but amazingly useful) principle: if $i$ and $j$ are itemsets such that $j$ is a subset of $i$ then the support of $j$ is greater than or equal to the support of $i$. Thus, for an itemset to be frequent all its subsets must in turn be frequent as well. This gives rise to an efficient level-wise construction of frequent itemsets in $D$. The algorithm makes multiple passes over the data. Starting with itemsets of size 1 (i.e. 1-itemsets), every pass discovers frequent itemsets of the next bigger size. The first pass over the data discovers all the frequent 1-itemsets. These are then combined to generate candidate 2-itemsets and by determining their supports (using a second pass over the data) the frequent 2-itemsets are found. Similarly, these frequent 2-itemsets are used to first obtain candidate 3-itemsets and then (using a third database pass) the frequent 3-itemsets are found, and so on. The candidate generation before the $m^{th}$ pass uses the Apriori principle described above as follows: an m-itemset is considered a candidate only if all $(m-1)$-itemsets contained in it have already been declared frequent in the previous step. As m increases, while the number of all possible m-itemsets grows exponentially, the number of frequent m-itemsets grows much slower, and as a matter of fact, starts decreasing after some m. Thus the candidate generation method in Apriori makes the algorithm efficient. This process of progressively building itemsets of the next bigger size is continued till a stage is reached when (for some size of itemsets) there are no frequent itemsets left to continue. This marks the end of the frequent itemset discovery process.

3. **Note Worthy Contribution in the field of proposed work :**

Mendes, L.F. Bolin Ding, Jiawei Han [21] introduces two effective methods for mining sequential patterns from data streams: the SS-BE method and the SS-MB method. The proposed methods break the stream into batches and only process each batch once. The two methods use different pruning strategies that restrict the memory usage but can still guarantee that all true sequential patterns are output at the end of any batch. Both algorithms scale linearly in execution time as the number of sequences grows, making them effective methods for sequential pattern mining in data streams. The experimental results also show that our methods are very accurate in that only a small fraction of the patterns that are output are false positives. Even for these false positives, SS-BE guarantees that their true support is above a pre-defined threshold.

Lei Chang Tengjiao Wang Dongqing Yang Hua Luan [22] studies the problem of mining closed sequential patterns over data stream sliding windows. An efficient algorithm SeqStream is developed to mine closed sequential patterns in stream windows incrementally, and various novel strategies are adopted in SeqStream to prune search space aggressively. Extensive experiments on both real and synthetic data sets show that SeqStream outperforms PrefixSpan, CloSpan and BIDE by a factor of about one to two orders of magnitude.

The input data is a set of sequences, called data-sequences. Each data sequence is a ordered list of transactions (or itemsets), where each transaction is a sets of items (literals). Typically there is a transaction-time associated with each transaction. A sequential pattern also consists of a list of sets of items. The problem is to find all sequential patterns with a user-specified minimum support, where the support of a sequential pattern is the percentage of data sequences that contain the pattern.

### 4. Proposed Methodology:

We have done study about pattern of different Result Analysis of Our University Result Data of Different Semesters as shown below.

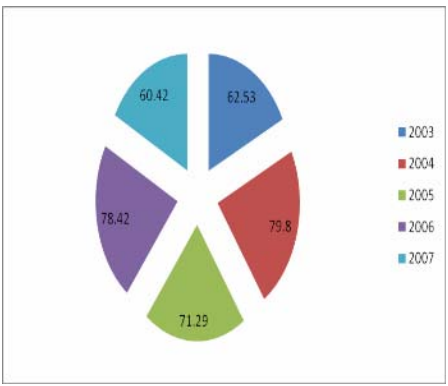

1.1 Result Graph for BE-101

Year wise Data for Subject Code BE-101

| Year | Rst _Per |
|------|----------|
| 2003 | 62.5 |
| 2004 | 79.8 |
| 2005 | 71.3 |
| 2006 | 78.4 |
| 2007 | 60.4 |

1.2 Year wise Data for Subject Code BE-101

| year | R_pst |
|------|-------|
| 2003 | 66.55 |
| 2004 | 68.69 |
| 2005 | 79.72 |
| 2006 | 72.66 |
| 2007 | 68.08 |

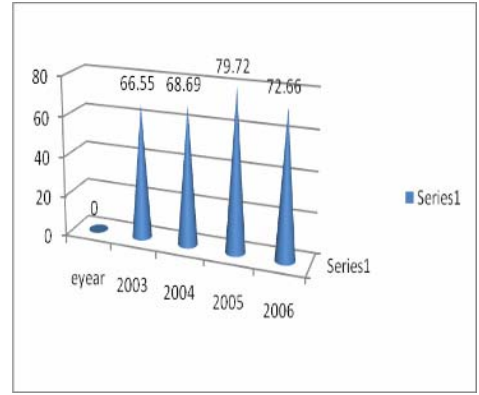

1.2 Result Graph for BE-102

1.3 Year wise Data for Subject Code BE-103

| Year | R_percent |
|------|-----------|
| 2003 | 88.62 |
| 2004 | 90.54 |
| 2005 | 91.57 |
| 2006 | 90.28 |
| 2007 | 90.94 |

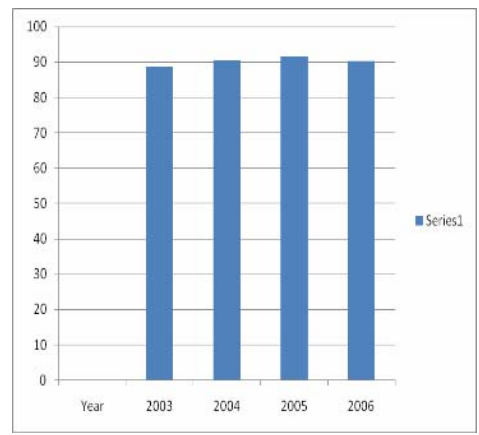

1.3 Result Graph for BE-103

1.4 Year wise Data for Subject Code BE-104

| year | R_pst |
|------|-------|
| 2003 | 88.62 |
| 2004 | 90.54 |
| 2005 | 91.57 |
| 2006 | 90.28 |
| 2007 | 90.94 |

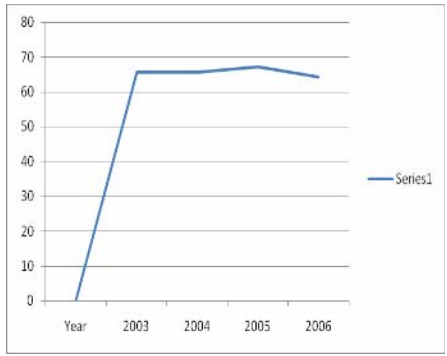

1.4 Result Graph for BE-104

1.5 Year wise Data for Subject Code BE-105

| Year | R_percent |
|------|-----------|
| 2003 | 72.8 |
| 2004 | 87.44 |
| 2005 | 69.45 |
| 2006 | 74.4 |
| 2007 | 29.69 |

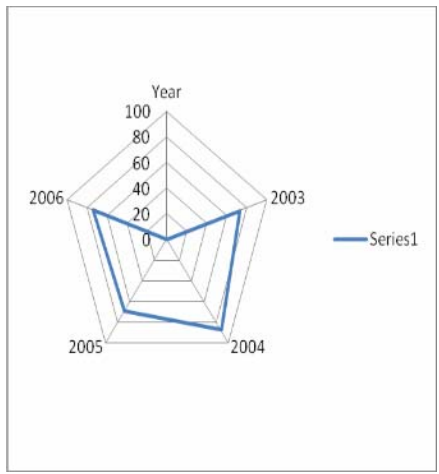

1.5 Result Graph for BE-105

The methodology applied to complete the research work of *"A Way to Undwerstand Various Pattern Mining Techniques for Selected Domain"* was divided into series of steps. We envisage to study and implement the following methods.

- Study of Temporal relations.
- Provide an overview, the research survey and summarizing previous work that investigated the various functions of data sequences in various domains.
- Problem formulation and generate the frequent sequences.
- Sub-division of the sequences based on structure of the sequence i.e. constraints based mining and extended sequence based mining with pruning strategies.
- Analysis and evaluation of the proposed sequential pattern mining algorithm with item gap and time stamp.

5. **Expected outcome of the proposed work :**

- Appropriate duration modeling for events in sequences.
- Improving time and space complexities of algorithms.
- Comparison with the existing models on extract sequence quality, number of extracted sequences and execution time.
- Implementation of the proposed sequential patterns.
- If implementation is successful then tested for evaluation.

6. **Bibliography in standard format :**